\documentclass[conference]{ifacconf}

\usepackage{cite}
\usepackage{amsmath,amssymb,amsfonts}
\usepackage{graphicx}\usepackage{natbib}\usepackage{tikz}\usepackage{svg}\usepackage{calc}
\usepackage{textcomp}

\usepackage{subfig}
\usepackage{mathrsfs}
\usepackage{algorithm}
\usepackage{wrapfig}

\usepackage{algpseudocode}
\usepackage{algorithm}
\def\NoNumber#1{{\def\alglinenumber##1{}\State #1}\addtocounter{ALG@line}{-2}}

\def\BibTeX{{\rm B\kern-.05em{\sc i\kern-.025em b}\kern-.08em
    T\kern-.1667em\lower.7ex\hbox{E}\kern-.125emX}}
\newcommand{\bld}[1]{\boldsymbol{#1}}
\DeclareMathOperator*{\argmin}{argmin}

\begin{document}
	\begin{frontmatter}
\title{Approximating a Laplacian Prior for Joint State and Model Estimation within an UKF\thanksref{footnoteinfo}} 

\thanks[footnoteinfo]{This work was developed in the junior research group DART (Daten\-ge\-trie\-be\-ne Methoden in der Regelungstechnik), Paderborn University, and funded by the Federal Ministry of Education and Research of Germany (BMBF - Bundesministerium für Bildung und Forschung) under the funding code 01IS20052. The responsibility for the content of this publication lies with the authors.\\ This work has been submitted to IFAC for possible publication.}

\author{Ricarda-Samantha Götte, Julia Timmermann} 

\address{Heinz Nixdorf Institute, Paderborn University, Paderborn, Germany
	(e-mail: \{rgoette,julia.timmermann\}@hni.upb.de)}

\begin{abstract}
A major challenge in state estimation with model-based observers are low-quality models that lack of relevant dynamics. We address this issue by simultaneously estimating the system's states and its model uncertainties by a square root UKF. Concretely, we extend the state by the parameter vector of a linear combination containing suitable functions that approximate the lacking dynamics. Presuming that only a few dynamical terms are relevant, the parameter vector is claimed to be sparse. In Bayesian setting, properties like sparsity are expressed by a prior distribution. One common choice for sparsity is a Laplace distribution. However, due to some disadvantages of a Laplacian prior, the regularized horseshoe distribution, a Gaussian that approximately features sparsity, is applied. Results exhibit small estimation errors with model improvements detected by an automated model reduction technique.  
\end{abstract}

\begin{keyword}
joint estimation, unscented Kalman filter, sparsity, Laplacian prior, regularized horseshoe, principal component analysis
\end{keyword}
\end{frontmatter}
\section{Introduction}
The concept of state estimation is central to control design since usually only a part of the system states can be measured but the complete state needs to be known for controlling the system sufficiently. Model-based observers offer accurate state estimates when a model covers the system's behavior closely. However, the state inference is much more challenging when model inaccuracies or uncertainties are present. 

Yet, one simple approach to approximate model uncertainties is the formulation as a linear combination of suitable mappings. Subsequently, the parameters of this linear combination need to be identified to deliver correct state estimates. As stated in \cite{Kullberg.2021} and \cite{Gotte.2022}, the states and model uncertainties can be jointly estimated by defining an extended state vector that contains the system's states along with the parameters. Accordingly, this state is then utilized within a filter, in this work a square root unscented Kalman filter (SRUKF) that employs the unscented transform (UT) (c.f. \cite{vanderMerwe.2001}). However, assumed that most dynamics in nature and technology are characterized by rather a few terms, the parameters are claimed to be sparse, leading to the objective of how to incorporate this into existing filter structures.  

While we addressed this challenge in our former work \cite{Gotte.2022} from an optimization perspective, we now take a stochastic view to demand sparsity for the linear combination's parameters and insert this claim within the SRUKF procedure. For this purpose, a prior distribution that expresses sparsity needs to be chosen. 
Previous research sometimes suggested the usage of a Laplacian distribution. However, e.g. \cite{Hirsh.2022} point out disadvantages of a Laplacian prior and recommend  using a specific Gaussian distribution, namely the regularized horseshoe distribution, that approximately features sparsity.       

Although state estimation is the primary goal, insight into model uncertainties to possibly improve and update the low-quality model might be a reasonable subordinate objective. Indeed, very few research in joint estimation has focused on the interpretation of the model inaccuracies. For instance, \cite{Kullberg.2021} deploy radial basis functions as suitable mappings but focus more on the no less important aspect of computational efficiency within state estimation. Likewise, \cite{Khajenejad.2021} present an approach to narrow the current state by geometrical limitations but do not extract these to derive an interpretation of the model inaccuracies. Conversely, as a by-product we apply a model reduction technique to automatically obtain the most dominant terms for model update.

Therefore, our main contributions are the following:
\begin{itemize}
	\item Joint estimation of states and model uncertainties within a SRUKF,
	\item efficient promotion of sparsity and exploitation of the filter structure by regularized horseshoe distribution, 
	\item automated extraction of interpretable terms for model update.
\end{itemize} 

The paper is organized as follows. In Sec. \ref{sec:problem} the joint model is defined and the claim of sparsity for the linear combination's parameters is motivated. Consequently, this claim is expressed by modeling the parameters with the regularized horseshoe distribution and deployed within an SRUKF in Sec. \ref{sec:method}, followed by experimental results in Sec. \ref{sec:Simulations}. Eventually, Sec. \ref{sec:interpretation} copes with the automated model update. The work concludes with a short discussion in Sec. \ref{sec:conclusion}.  

\textit{Notation:} $\bld{\tilde{\bullet}}$ denotes the joint state vector or relating variables, such as the covariance matrix. Bold variables refer to vectors or matrices, others to scalars. A subscript $i$ denotes the $i$th element of a vector, while the subscript $k$ indicates the current time step. $\hat{\bullet}$ describes estimated variables.

\section{Problem Formulation}\label{sec:problem}
Due to the complexity of a system, modeling errors or lack of time, one often is forced to apply a low-quality model $\bld{x}_{k+1}=\bld{f}(\bld{x}_k,u_k)$ to estimate the states with a model-based observer. Clearly, this may lead to an insufficient estimation quality. However, based on our previous work \cite{Gotte.2022} we assume the unknown partial dynamics $\bld{g}(\bld{x}_k,u_k)$ as a mapping that can be approximated by linear combinations of $n_{\bld{\theta}}$ suitable functions $\psi_i(\bld{x},u)$, stored in a library $\bld{\Psi}$. Those depend on the state $\bld{x}_k\in\mathbb{R}^{n_{\bld{x}}}$ as well as on the control input $u_k\in\mathbb{R}$ at time step $k$ and are evaluated by parameters $\bld{\theta}_k\in\mathbb{R}^{n_{\bld{\theta}}}$. Thus, the system's dynamics are characterized by a joint state $\tilde{\bld{x}}_k = \left(\bld{x}_k^T,\bld{\theta}_k^T\right)^T\in\mathbb{R}^{\tilde{n}}$ with $\tilde{n}=n_{\bld{x}}+n_{\bld{\theta}}$ as follows:
\begin{align}\label{eq:JEmodel}
\begin{split}
\tilde{\bld{x}}_{k+1}&=\begin{pmatrix}
\bld{f}(\bld{x}_k,u_k,\bld{g}(\bld{x}_k,u_k))+\bld{w}_k^{\bld{x}}\\
\bld{\theta}_k+\bld{w}_k^{\bld{\theta}}
\end{pmatrix},\\
&=\begin{pmatrix}
\bld{f}(\bld{x}_k,u_k,\bld{\theta}_k^T\bld{\Psi}(\bld{x}_k,u_k))+\bld{w}_k^{\bld{x}}\\
\bld{\theta}_k+\bld{w}_k^{\bld{\theta}}
\end{pmatrix}\\&=:\tilde{\bld{f}}(\tilde{\bld{x}}_k,u_k)+\bld{w}_k,\\
\bld{y}_k&=\bld{h}({\bld{x}}_k,u_k)+\bld{v}_k.	
\end{split}
\end{align}
Here, $\bld{f}$ and $\bld{h}$ denote the dynamical and observational models, respectively, while $\tilde{\bld{f}}$ refers to the joint model equipped with $\bld{\theta}_k$. The system $(\bld{f},\bld{h})$ is assumed to be observable by measurements $\bld{y}_k\in\mathbb{R}^m$. Note that $\bld{h}$ is evaluated only for $\bld{x}_k$ since the parameters can not be measured. Further, remark that $\bld{\theta}_k$'s evolution is modeled by stationary dynamics and additive Gaussian noise $\bld{w}_k^{\bld{\theta}}\sim\mathcal{N}(\bld{0},\bld{Q}_{\bld{\theta}})$, whereas it holds $\bld{w}_k^{\bld{x}}\sim\mathcal{N}(\bld{0},\bld{Q}_{\bld{x}})$ and $\bld{v}_k\sim\mathcal{N}(\bld{0},\bld{R})$. However, $\bld{\theta}_k$'s values are initially unknown and may cause an unstable or divergent observer without further assumptions due to the numerous degrees of freedom if the quotient $n_{\bld{\theta}}/n_{\bld{x}}$ tends towards zero (c.f. \cite{Gotte.2022}). Hence, presuming that only a small part of $\bld{\Psi}$ is indeed present in the system's dynamics, the parameters $\bld{\theta}_k$ are expected to be sparse. Thus, how to incorporate this claim within established observers, such as the SRUKF, needs to be addressed.     

In our previous work we aimed at this question from an optimization perspective. To receive a small estimation error but also ensure $\bld{\theta}_k$ remains sparse, the solution of the following multi-objective problem is sought for:
\begin{equation}\label{eq:filterProblem}
\argmin_{\bld{\hat{\tilde{x}}}_k} \frac{1}{2}\mathbb{E}[(\bld{\hat{\tilde{x}}}_k-\bld{\tilde{x}}_k)^T(\bld{\hat{\tilde{x}}}_k-\bld{\tilde{x}}_k)]+\lambda \vert\vert \bld{\tilde{I}}\bld{\hat{\tilde{x}}}_k\vert\vert_1
\end{equation}
with $\bld{\tilde{I}}=\text{blkdiag}(\bld{0}_{n_{\bld{x}}},\bld{I}_{n_{\bld{\theta}}})$.
By reformulating \eqref{eq:filterProblem} into a minimization problem with a soft constraint that is used as a pseudo measurement, it could be proven that correct estimates were delivered, while also interpretable results for the model updates were discovered. However, this strategy may become elaborate as it requires multiple performance parameters to be preset by the user and increases its complexity when higher dimensional model inaccuracies arise. 

Since the model \eqref{eq:JEmodel} should be used within a Bayesian filter, it is only natural to consider a stochastic viewpoint for the claim of sparsity and derive a more efficient strategy.

\section{Joint estimation via an approximated Laplacian prior}\label{sec:method}
In classical Bayesian filters the state $\bld{x}_k$, its evolution and observation are assumed to be Gaussian distributed. Claiming that $\bld{\theta}_k$ needs to be sparse, the question arises which distribution captures sparsity most naturally. One of the most prominent distributions, that induces sparsity, is the Laplace distribution.

\subsection{Laplace distribution and sparsity}
Since the Lasso method is equivalent to Bayesian regression with a Laplacian prior (\cite{Hastie.2015}), it is only intuitive to consider this distribution for promoting sparsity within our setting. Its probability density function for a single $\theta_i$ is defined by
\begin{equation}\label{eq:Laplace_pdf}
	p(\theta_i \vert \mu,\sigma)=\frac{1}{2\sigma}\exp\left(-\frac{\vert \theta_i-\mu\vert}{\sigma}\right)
\end{equation}
with parameters $\mu\in\mathbb{R},\sigma\in\mathbb{R}^+$.
Admittedly, the usage of a Laplacian prior is not intended within a SRUKF. Functionality is only provided if the underlying distribution is Gaussian (\cite{vanderMerwe.2001}). Yet, \cite{Ebeigbe.2021} present a generalized version of the unscented transform to apply other distributions than Gaussians like the Laplacian in Eq. \eqref{eq:Laplace_pdf}. 
   
Nonetheless, as \cite{Hastie.2015} and \cite{Hirsh.2022} point out using a Laplacian prior does not result in a posterior distribution that keeps the Laplacian shape. \cite{Hirsh.2022} further discuss an example that illustrates the wide range of a resulting posterior distribution when a Laplacian prior is applied. Even though the means are mostly centered around the true coefficients, it is challenging to distinguish between dominant and non-dominant coefficients due to the distributed probability mass. Therefore, the Laplacian is an inappropriate prior when estimating model uncertainties and keeping $\bld{\theta}$ sparse.
Instead, it is more practical to employ a Gaussian that imitates and keeps the shape of a Laplacian by adjusting and changing its variance. By this, it is still applicable to the classical SRUKF without almost any changes regarding weights or performance parameters necessary.

\begin{figure}[h]
	\centering
	\def\svgwidth{\columnwidth}
	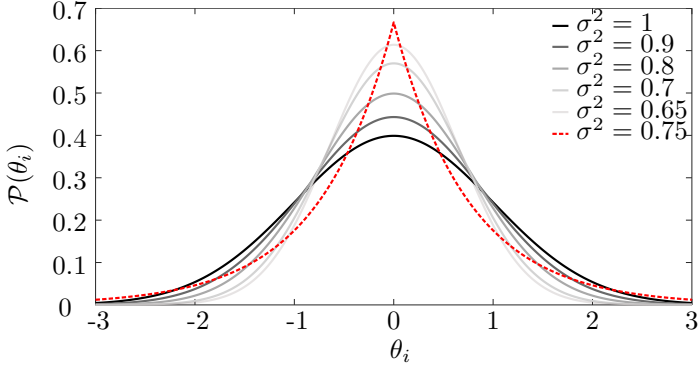
	\caption{Forming a standard Gaussian (black solid) towards a Laplacian shape (red dashed) by adjusting its variance (gray solid).}\label{fig:LaplaceGauss}
\end{figure}

In Fig. \ref{fig:LaplaceGauss} the idea of imitating the Laplacian shape by a Gaussian is sketched. By shrinking its variance $\sigma^2$ from standard to $0.65$ (ranging from black to light gray), the Gaussian keeps getting closer to the shape of this exemplary chosen Laplacian with variance of $0.75$, displayed in red dashed. Now, how to chose the variance explicitly to force the Gaussian towards a sparse shape, remains an open design question.
Previous works, e.g. \cite{Piironen.2017}, \cite{Bhadra.2019} and \cite{Hirsh.2022}, have shown that among other Gaussians distributions the regularized horseshoe distribution stands out as a very suitable choice for approximating the Laplacian.

\subsection{Regularized horseshoe prior within a SRUKF}
As pointed out earlier, a Gaussian distribution with a variance, that is defined over distributions itself, manages to imitate the desired Laplacian shape while remaining applicable to a classical Bayesian filter. In this work the regularized horseshoe distribution is used and characterized by the following for a single $\theta_i$:
\begin{align}\label{eq:regularizedHorseshoe}
\begin{split}
\theta_i\vert{\lambda}_i,\tau,c&\sim\mathcal{N}(0,\check{\lambda}_i^2\tau^2),\\
\lambda_i&\sim\mathcal{C}^+(0,1),\\
\tau&\sim\mathcal{C}^+(0,\tau_0),\\
c^2&\sim\text{Inv-}\Gamma(a,b),\\
\check{\lambda}_i &= \frac{c\lambda_i}{\sqrt{c^2+\tau^2\lambda_i^2}}.
\end{split}
\end{align}
Obviously, the $i$th variance $\sigma_i^2=\check{\lambda}_i^2\tau^2$ is derived by the product of two half-Cauchy distributions $\mathcal{C}^+$ with a regularization parameter $c^2$ that itself is inverse Gamma distributed $\text{Inv}-\Gamma$ with coefficients $a,b \in \mathbb{R}^+$. 
While $\tau$ pushes the parameters $\bld{\theta}$ to be globally sparse by reducing the posterior distribution, the local shrinkage parameter $\lambda_i$ enables some $\theta_i$ to elude the global sparsity. Hence, decreasing the hyperparameter $\tau_0$ leads to a stronger global sparsity and therefore to more sparse $\bld{\theta}$ estimates (c.f. \cite{Hirsh.2022}, \cite{Piironen.2017}).

Thus, the joint model \eqref{eq:JEmodel} is assumed to comprise of two Gaussians, namely $\bld{x}_k\sim\mathcal{N}(\bld{0},\bld{I})$ and ${\theta}_{i,k}\vert\lambda_i,\tau,c\sim\mathcal{N}({0},{\check{\lambda}}_i^2\tau^2)$. For simplicity, one $\sigma^2_i$ is assumed for all $\theta_i$. As mentioned earlier, this strategy is almost directly applicable within the SRUKF. For details regarding the SRUKF algorithm, we refer to \cite{vanderMerwe.2001} and for notation issues to \cite{Gotte.2022}. Due to the approximation of the fourth moment for a Gaussian distribution it exists an optimal value for the performance parameter $\kappa$ (c.f. \cite{Julier.1995} and \cite{Julier.1997}). Unfortunately, since herein two different Gaussian distributions are considered, the advantage of modeling sparsity by a specific Gaussian results in the need for two different $\kappa$'s. For  a standard Gaussian random variable $\mathcal{Z}_1$ it holds $\mathbb{E}[\mathcal{Z}_1^4]=3$, whereas for an arbitrary Gaussian with a random variable $\mathcal{Z}_2$ it holds $\mathbb{E}[\mathcal{Z}_2^4]=3\sigma^2_2$. Although we have
\begin{align}\label{eq:twoGaussian}
\begin{split}
\mathbb{E}[(\mathcal{Z}_2-\hat{z}_2)^4]&=\mathbb{E}[((\hat{z}_2+\sqrt{P_z}\mathcal{Z}_1)-\hat{z}_2)^4]\\&=\mathbb{E}[(\sqrt{P_z}\mathcal{Z}_1))^4]
\end{split}
\end{align} 
with covariance matrix $P_z$, this usually does not represent the true fourth moments of $\mathcal{Z}_2$ since it is only optimized for $\mathbb{E}[\mathcal{Z}_1^4]$.   
Hence, a two-step approach, one for each Gaussian, is proposed and outlined via pseudo code in Algo. \ref{algo:pseudocode} and the following explanations. Alternatively, a $\kappa$ could be found by optimization (c.f. \cite{Schweers.2013}, and \cite{Turner.2012}) or by analytically deriving a $\kappa$ that provides a compromise by minimizing the fourth moment errors of both distributions equally well.

\begin{algorithm}
	\caption{Joint SRUKF with stochastic perspective}
	\begin{algorithmic}[1]
		\Require $\tilde{\bld{x}}_0,\bld{S}_0,\alpha,\beta,\tilde{\bld{Q}},\bld{R},R_{pm},\tau_0,a,b$
		\For{$k=0,\dots$}
		\State $\bld{W}_m^{(1)},\bld{W}_c^{(1)}\gets\alpha,\beta,\kappa^{(1)}=3-\tilde{n}$
		\State $\tilde{\bld{x}}^{(1)}_{k+1\vert k},\bld{S}^{(1)}_{k+1\vert k} \gets \text{SRUKF with}\,\bld{\tilde{f}}, \,\bld{h},\,\tilde{\bld{Q}},\,\bld{R}$
		\NoNumber	\State
		
		\State $\sigma^2_{\star}=\mathbb{E}[\sigma^2]\gets\,\text{Determine by Eq. \eqref{eq:regularizedHorseshoe}\,with\,}  \tau_0,a,b$
		\State $\bld{W}_m^{(2)},\bld{W}_c^{(2)}\gets\alpha,\beta,\kappa^{(2)}=3\sigma^2_{\star}-\tilde{n}$
		\State $\tilde{\bld{x}}^{(2)}_{k+1\vert k},\bld{S}^{(2)}_{k+1\vert k} \gets \text{SRUKF with}\, \bld{f}_{Id},\,{h}_{pm},\,\tilde{\bld{Q}},\,R_{pm} $
		\NoNumber	\State
		
		\State $\tilde{\bld{x}}^-_{k+1\vert k}\gets \text{Merge}\,\tilde{\bld{x}}^{(1)}_{k+1\vert k}\,\text{and}\,\tilde{\bld{x}}^{(2)}_{k+1\vert k}$
		\State $\bld{S}^-_{k+1\vert k}\gets \text{Merge}\,\bld{S}^{(1)}_{k+1\vert k}\,\text{and}\,\bld{S}^{(2)}_{k+1\vert k}$
		\EndFor
	\end{algorithmic}\label{algo:pseudocode}
\end{algorithm}

Initially, a SRUKF routine is run through for $\kappa^{(1)}=3-\tilde{n}$, assuming a standard Gaussian. Here, the joint model \eqref{eq:JEmodel} is utilized within the UT as dynamical and observational model (lines 2 and 3 in Algo. \ref{algo:pseudocode}). After that, a second SRUKF routine is enclosed that determines the variance $\sigma^2_{\star}$ and employs $\kappa^{(2)}=3\sigma^2_{\star}-\tilde{n}$.  In contrast to the first SRUKF routine, the dynamical model is now the identity $\bld{f}_{Id}$, as only the measurements need to be adjusted. They are derived by a pseudo observational model $h_{pm}$ that quantifies the sparsity of $\bld{\theta}_k$ by
\begin{equation}\label{eq:pseudomodel}
	y_{pm}=h_{pm}(\tilde{\bld{x}})=\max\{\vert\vert\bld{\tilde{I}}\bld{x}\vert\Vert_1-\epsilon,0\}.
\end{equation}
The pseudo measurement is a strategy we already utilized in our work \cite{Gotte.2022} and has been proposed by \cite{Carmi.2010}. Since Eq. \eqref{eq:pseudomodel} is employed as observational model, it also holds a pseudo measurement covariance due to its noise $\epsilon>0$, that is denoted by $R_{pm}$ and usually initialized with a large number (lines 4 to 6). Consequently, the posterior estimates after the first routine, indicated by superscripts $\bullet^{(1)}$, need to be adjusted. Their $\bld{\theta}$-related blocks are replaced by posterior estimates of the second routine, denoted by superscripts $\bullet^{(2)}$. Specifically, this means the (quadratic) blocks from $n_{\bld{x}}+1$ up to $n_{\bld{x}}+n_{\bld{\theta}}$. Ultimately, this procedure delivers the estimated posterior mean and covariance (lines 7 and 8).

The remaining SRUKF's performance parameters, that determine the sigma points' position compared to the mean, are chosen for both routines with $\alpha=0.001$ and $\beta=2$ since both are assumed to be Gaussian distributed. The covariance matrices $\tilde{\bld{Q}}$, that consists of the block matrices $\bld{Q}_{\bld{x}}$ and $\bld{Q}_{\bld{\theta}}$ with zeros elsewhere, and  $\bld{R}$ are initialized with small numbers. For the regularized horseshoe distribution it holds $\tau_0=0.1, a=4.5$ and $b=1.5$ throughout the paper.
\subsection{Observability}
Since the classical SRUKF's procedure is adjusted, we investigate the proposed algorithm's stability and related properties. Therefore, the preceding question is the prerequisite of observability. Since the observational model $\bld{h}$ does not allow to infer the linear combination's parameters $\bld{\theta}$, the joint model is first not observable though the system $(\bld{f},\bld{h})$ itself is observable. However, by the introduction of a pseudo measurement $h_{pm}$ the parameters $\bld{\theta}$ become observable. Merging both observational models to $\tilde{\bld{h}}=(\bld{h},h_{pm})$, we are able to investigate the observability of the joint system $(\tilde{\bld{f}},\tilde{\bld{h}})$. Since the parameters $\bld{\theta}$ will always differ from zero during the numerical filter calculations if not initialized with zero, the function $h_{pm}$ remains differentiable. Thus, we derive the following observability mapping $\bld{\Phi}$ that contains the $l$th Lie derivative $L_{\tilde{\bld{f}}}^{l}\tilde{\bld{h}}_j(\tilde{\bld{x}},u)$ with respect to $\tilde{\bld{f}}$ and ${l=0,\dots,\tilde{n}-1,}\,j=1,\dots,m+1$:
\begin{equation}\label{eq:observabilitymapping}
	\bld{\Phi}(\bld{\tilde{x}},u)=\begin{pmatrix}
	L_{\tilde{\bld{f}}}^{0}\tilde{\bld{h}}_1(\tilde{\bld{x}},u)\\\vdots\\
	L_{\tilde{\bld{f}}}^{0}\tilde{\bld{h}}_{m+1}(\tilde{\bld{x}},u)\\
	\vdots\\
	L_{\tilde{\bld{f}}}^{\tilde{n}-1}\tilde{\bld{h}}_1(\tilde{\bld{x}},u)\\
	\vdots\\
	L_{\tilde{\bld{f}}}^{\tilde{n}-1}\tilde{\bld{h}}_{m+1}(\tilde{\bld{x}},u)
	\end{pmatrix}.
\end{equation}
For brevity, the subscript $k$ denoting the current time step has been omitted. Since the observability matrix $\bld{Q}_{obs}\in\mathbb{R}^{\tilde{n}(m+1)\times \tilde{n}}$, defined by 
\begin{equation}\label{eq:observability matrix}
\begin{split}
	\bld{Q}_{obs}=\frac{\partial \bld{\Phi}}{\partial \tilde{\bld{x}}},
\end{split}
\end{equation} 
has full column rank $\tilde{n}$ for $\bld{x}\in\mathbb{R},{\theta}_i\in\mathbb{R}\backslash\{0\}$, the system is observable. Due to $(\bld{f},\bld{h})$ being observable and $\bld{\theta}$ measured by $h_{pm}$, we are able to draw conclusions for the system $(\tilde{\bld{f}},\tilde{\bld{h}})$. Additionally, this holds, even if a two-step approach in Algo. \ref{algo:pseudocode} is pursued. Stability of the proposed filter algorithm is given due to SRUKF's properties.

\section{Experimental results}\label{sec:Simulations}
This section deals with the deployment and test of the aforementioned method. Results on rather a simple example are discussed to effectively comprehend and demonstrate the method and its effects in depth. Other applications have shown to confirm these observations. Since the primary aim is accurate state estimation and the subordinate goal is the detection of model uncertainties, we focus on both separately.

\subsection{State estimation}
Initially, we consider the simple system of the Duffing oscillator that has been described within our previous work \cite{Gotte.2022}. Its dynamics 
\begin{align}\label{eq:Duffing}
\begin{split}
\bld{\dot{x}}&=\begin{pmatrix}
x_2\\ -p_3x_2-p_1x_1-p_2x_1^3+u
\end{pmatrix},\\
y&=x_1,
\end{split}
\end{align}
with parameters $\bld{p}=(-1,3,0.1)^T$  are advantageous to investigate because we can define $g(\bld{x},u)=-p_2x_1^3$ as the lacking dynamical term but are able to validate the approach's correctness. Thus, the joint model is formulated by
\begin{align}\label{eq:DuffingJE}
	\begin{split}
	{\tilde{\bld{x}}}_{k+1}&=\begin{pmatrix}
	x_{2,k}+{w}_k^{x_1}\\ -p_3x_{2,k}-p_1x_{1,k}+\bld{\theta}_k^T\bld{\Psi}(\bld{x}_k,u_k)+u_k+{w}_k^{x_2}\\\bld{\theta}_k+\bld{w}_k^{\bld{\theta}}
	\end{pmatrix},\\
	y&=x_{1,k}.
	\end{split}
\end{align}  
Using $\bld{\Psi}(\bld{x},u)=(1,x_1,x_2,x_2^2,\sin(x_2),x_1^2,x_1x_2,\cos(x_1),u)^T$ $\in\mathbb{R}^9$, the model \eqref{eq:DuffingJE} is evaluated by explicit Euler with time step $\Delta t$. Clearly, the library $\bld{\Psi}$ does not contain the true term $g$. Indeed, it is the more interesting and realistic case to consider hypotheses that may be close to the true system's behavior. However, experiments have proved that if the true lacking term is included it will be selected with a high probability (c.f. \cite{Gotte.2022}). 

\begin{figure}[h]
	\centering

	\def\svgwidth{0.95\columnwidth}
	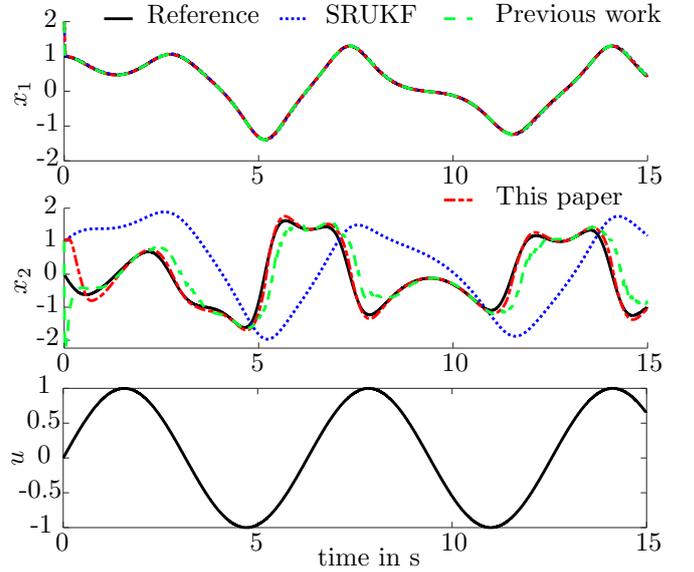
	\caption{Compared state estimation of a reference signal (solid black) by a classical SRUKF with the low-quality model (blue dotted), our previous optimization based approach (green dashed) and the proposed method (red dashed).}\label{fig:Duffing_x_Psi3}
\end{figure}

In Fig. \ref{fig:Duffing_x_Psi3} the estimated states for a sinusoidal signal are displayed compared to the reference trajectory in black. The true initial state is $\bld{x}_0=(1,0)^T$, whereas the observers start with the corrupted state $\bld{\hat{x}}_0=(2,1)^T$ and the parameters $\bld{\hat{\theta}}_0$ are initialized with small values. Two observers that utilize joint models are compared to the reference trajectory provided in solid black: one based on an optimization view from our previous work (in green dashed) and the proposed observer within this work relying on a stochastic perspective (in red dashed). Further, the comparison also features a classical SRUKF that employs the corrupted model without $g$ and is depicted in blue dotted.

\begin{wrapfigure}[]{r}{0.5\columnwidth}
	\centering

	\def\svgwidth{0.4\columnwidth}
	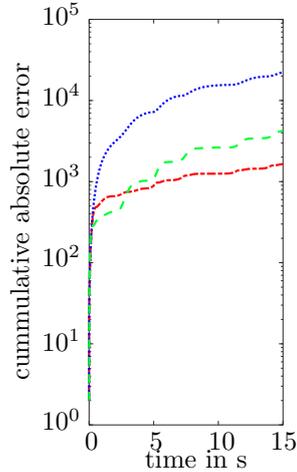
	\caption{Cumulative error to quantify the observers' accuracy.}\label{fig:Duffing_x_Psi3_error}
\end{wrapfigure}

Both observers based on joint models outperform the classical observer substantially that is not able to estimate the correct trajectory. Both also provide a fast reduction of the initial error that is proved by the close estimate for $x_2$ at approximately 1$s$. However, the proposed observer within this work appears to be slightly more accurate. This impression is verified by the cumulative error over time in Fig. \ref{fig:Duffing_x_Psi3_error}. In summary, the herein proposed observer manages to deliver accurate state estimates reliably although the true model inaccuracy $g$ is not included in $\bld{\Psi}$. Nonetheless, it is noteworthy to emphasize that regarding state estimation the proposed observer is not much more accurate than the one from previous work. However, the next section will reveal substantial differences between them and highlight the advantages of the former one concerning the interpretation of the model inaccuracies.

\subsection{Estimation of the model inaccuracies}
Since the primary goal of state estimation has been discussed previously, this section deals with the subordinate aim of estimating model inaccuracies.

\begin{figure}[h]
	\centering

	\def\svgwidth{0.95\columnwidth}
	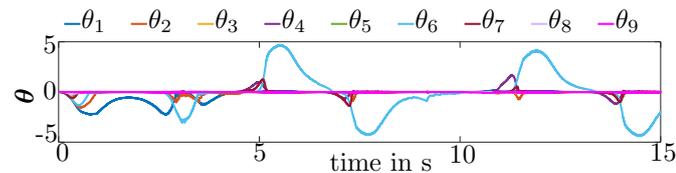
	\caption{Evolution of $\bld{\theta}$ over time exhibits that, besides $\psi_1(\bld{x},u)=1$, $\psi_6(\bld{x},u)=x_1^2$ is the most present term and reflects the correct sign of $g$ by its oscillating parameter $\theta_6$.}\label{fig:Duffing_Psi3}
\end{figure}

\begin{figure}[h]
	\centering

	\def\svgwidth{\columnwidth}
	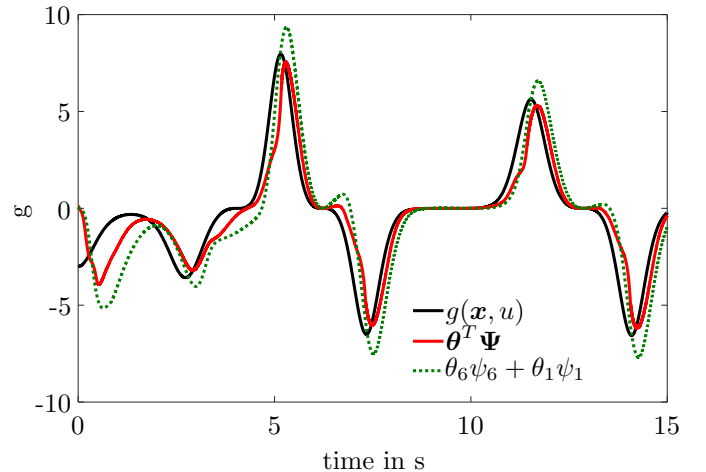
	\caption{Evolution of true model inaccuracy $g$ (black line) and its corresponding approximation by $\bld{\theta}^T\bld{\Psi}$ (red line) are compared. The most dominant terms identified by PCA (see Sec. \ref{sec:interpretation}) approximate $g$ by the green dotted line.}\label{fig:Duffing_Psi3_g}
\end{figure}

In Fig. \ref{fig:Duffing_Psi3} the evolution of parameters $\bld{\theta}$ corresponding to the states in Fig. \ref{fig:Duffing_x_Psi3} is depicted. Most obviously, the filter finds dominant $\{\psi_i\}_i$ that describe the model inaccuracy $g(\bld{x},u)=-p_2x_1^3$ best. In this case it is visually clear that the light blue and the dark blue lines are the most present ones, referring to $\psi_6(\bld{x},u)=x_1^2$, $\psi_1(\bld{x},u)=1$ respectively. Since $\bld{\Psi}$ does not contain the true $g$, the filter seeks for an alternative. Interestingly, the negative sign of $g$ gets maintained by switching the sign of $\theta_6$ for $\psi_6$. How close this alternative captures the actual inaccuracy is displayed in Fig. \ref{fig:Duffing_Psi3_g}. These results suggest that using a joint model with encoding sparsity via a regularized horseshoe distribution and a pseudo measurement allows not only accurate state estimation but also a deeper insight into model deviations, even if the true mathematical formulation can not be found. As suitable alternatives are provided we claim that more insight is always preferable in contrast to no understanding. Moreover, these insights could be processed and deepened by adjusting the library $\bld{\Psi}$, removing disproved hypotheses while adding new ones.

Compared to our former approach in \cite{Gotte.2022} the distinction between dominant and non-dominant terms is now carried out automatically without any user inputs. Thus, we do not need to preset any number for allowed non-sparse parameters or any limits to which a parameter is considered as sparse any more. Further, the $\bld{\theta}$'s smoother course over time clearly stands out to the former approach and can be reasoned with the efficient modeling and formulation of $\bld{\theta}$ as a distribution. Thus, no extra handling for the parameters' update like the soft switching approach is required any longer. However, if scaling becomes necessary it is easy to include it within this formulation by a constant $\xi_i$ in $\theta_i\sim\mathcal{N}(0,\check{\lambda}_i^2\tau^2)\xi_i$.     

\section{Extraction of interpretable terms}\label{sec:interpretation}
Since estimating the system's states correctly is the primary aim, the temporary approximation of its model uncertainties $\bld{g}(\bld{x}_k,u_k)$ as shown in Fig. \ref{fig:Duffing_Psi3} by the linear combination $\bld{\theta}_k^T\bld{\Psi}(\bld{x}_k,u_k)$ is the necessary means to an end. However, we can also use this by-product to update and improve the system's model. During time steps $k=1,\dots,N$, the observer collects data $\bld{\theta}_k$, that can be saved and used to analyze which dynamical terms $\psi_i$ have shown to be most dominant. Therefore, the samples $\bld{\theta}_k$ are stored column-wise within a matrix $\bld{\Theta}\in\mathbb{R}^{n_{\bld{\theta}}\times N}$. Still assuming that most dynamics encountered in control engineering can be characterized rather by a small amount of terms, model reduction techniques can be applied towards $\bld{\Theta}$. Within this work, we utilize one of the most popular ones, called principal component analysis (PCA) that \cite{Pearson.1901} and \cite{Hotelling.1933} originally introduced in early last century. This method seeks the most dominant features within high dimensional data and is commonly used in data science, e.g. \cite{Brunton.2019}.
\begin{figure}[h]
	\centering
	\def\svgwidth{\columnwidth}
	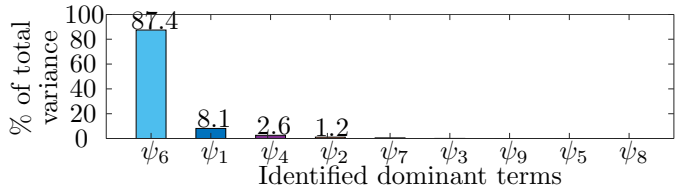
	\caption{PCA identifies dominant terms $\psi_i$ by analyzing $\bld{\Theta}$'s eigenvalues and their dispersion (displayed in percentage). }\label{fig:Duffing_Psi3_PCA}
\end{figure}

Deploying PCA on results from the former experiment in Sec. \ref{sec:Simulations}, the visual and qualitative impression of Fig. \ref{fig:Duffing_Psi3} is confirmed by a quantitative result. In Fig. \ref{fig:Duffing_Psi3_PCA} the percentage of the total variance correlates to the terms that are most dominant during the considered period. In this case $\psi_6(\bld{x},u)=x_1^2$ is the most dominant term with around $87\%$, followed by constants. Both capture over $95\%$ of the model inaccuracy $g$ which is a confident level to be accepted for a model update. The resulting $\hat{g}=\theta_6\psi_6+\theta_1\psi_1$ is displayed in Fig. \ref{fig:Duffing_Psi3_g} as green dotted and reveals a close approximation for the majority of the model inaccuracy $g$. Besides, it should be noted that it is often more meaningful to apply PCA after the transient behavior, herein after the first two seconds, leading to more intelligible insights. Ultimately, model update is then possible by including the identified dominant terms and applying a classical parameter identification scheme or a joint parameter approach.  
In general, the PCA study is conducted offline since a reasonable amount of samples is necessary to extract the information. Yet, a moving horizon is imaginable that allows to carry on the analysis during online estimation.

\section{Conclusion and outlook}\label{sec:conclusion}
To allow accurate state estimation when model inaccuracies occur, a method that encodes sparsity in a Bayesian filter has been proposed. In contrast to former works, the parameters of the linear combination, that approximates the model inaccuracies, are modeled via a Gaussian distribution whose variance is distributed as well to imitate a Laplacian shape. 
As a by-product the interpretation of these model deviations has been developed by an automatic model reduction technique. Illustrated on a simple example, the efficiency and sustainable benefit of the proposed method regarding a deeper system insight has been showed. Yet, future research will focus on the scheme's extension for higher dimensional $\bld{g}$ and on the test within a closed loop, including the need for online model updates.

\end{document}